\documentclass{aastex701}

\usepackage{adjustbox}
\usepackage{rotating}
\usepackage{amsmath}
\usepackage{makecell}
\usepackage{multirow}
\usepackage{booktabs}

\hypersetup{linkcolor=red,citecolor=blue,filecolor=cyan,urlcolor=magenta}

\begin{document}

\title{BSN: Light Curve Modeling and Orbital Evolution of the Total-Eclipse Contact Binary EZ Oct}

\author[0009-0004-8579-7692]{Asma Ababafi}
\affiliation{BSN project; Independent researcher, Khuzestan, 64615 Dezful, Iran}
\email{asma.ababafi@gmail.com}

\author[0000-0002-0196-9732]{Atila Poro}
\altaffiliation{atilaporo@bsnp.info, atila.poro@obspm.fr}
\affiliation{LUX, Observatoire de Paris, CNRS, PSL, 61 Avenue de l'Observatoire, 75014 Paris, France}
\affiliation{Astronomy Department of the Raderon AI Lab., BC., Burnaby, Canada}
\email{atila.poro@obspm.fr}

\author[0000-0002-3263-9680]{Mehmet Tanriver}
\affiliation{Department of Astronomy and Space Science, Faculty of Science, Erciyes University, Kayseri TR-38039, Türkiye}
\affiliation{Erciyes University, Astronomy and Space Science Observatory Application and Research Center, Kayseri TR-38039, Türkiye}
\email{mtanriver1@gmail.com}

\author[0000-0002-9262-4456]{Eduardo Fernández Lajús}
\affiliation{Instituto de Astrofísica de La Plata (CCT La Plata-CONICET-UNLP), La Plata, Argentina}
\email{eflajus@fcaglp.unlp.edu.ar}

\begin{abstract}
We present the first detailed multiband ($BVR_cI_c$ and TESS) photometric analysis of the short-period binary EZ Oct. This study combines ground-based observations conducted at a Southern Hemisphere observatory in Argentina with data from the TESS mission. Investigating the orbital period variations of EZ Oct reveals a steadily increasing period consistent with a quadratic trend. We present a new ephemeris and estimate the mass transfer rate as $\dot{M}=1.353\times 10^{-8}$ $M_{\odot}$/year, indicating ongoing conservative mass transfer from the less massive to the more massive star. Light curve modeling was performed using the PHOEBE Python code in conjunction with the MCMC approach, and the inclusion of a cold starspot was required to achieve an adequate fit. Absolute parameters were estimated using Gaia DR3 parallax and astrophysical equations. Our analysis shows that EZ Oct is a total-eclipse contact binary with a mass ratio of 1.969, a fillout factor of 0.106, and an inclination of $82.13^\circ$. Based on the stellar masses and temperatures of the components, the target system belongs to the W-subtype of contact binaries. The positions of the component stars were displayed on the mass–luminosity and mass–radius diagrams to illustrate their evolutionary status. Moreover, we investigated the relationship between orbital period and stellar luminosity in contact binary stars using a sample of 461 systems with $P < 0.5$ days. We highlight the position of EZ Oct in the mass ratio–inclination parameter space, showing that it lies within the densely populated region of contact binaries.
\end{abstract}


\section{Introduction}
Eclipsing binary systems play a pivotal role in astrophysics, offering essential insights into star formation, internal stellar structure, and the physical as well as evolutionary properties of stars (e.g., \citealt{2012ocpd.conf...51S}; \citealt{2014NewAR..60....1S}). Understanding these processes in W UMa-type binaries is crucial for constraining theoretical models of stellar evolution and for interpreting the statistical properties of observed close binary populations.

Based on Roche geometry, interacting binaries are typically grouped by the extent of stellar filling: systems in which both stars within their Roche lobes, systems where one component fills the Roche lobe, and contact systems in which both components share a common envelope (\citealt{1959cbs..book.....K}). In W Ursae Majoris (W UMa)-type binaries, both stars overfill their Roche lobes and exchange mass and energy efficiently (\citealt{van1982evolutionary}). When a component reaches its Roche-lobe limit, the gravitational potential promotes the flow of material and energy between the stars (\citealt{1971ARAA...9..183P, 2005ApJ...629.1055Y, 2006MNRAS.368.1311P}).

Modern studies show that this mass and energy exchange is tightly connected to orbital-period variations, mass-transfer episodes, and angular momentum loss (AML) processes, all of which significantly influence the long-term evolution of contact binaries. These variations are often interpreted within the Thermal Relaxation Oscillation (TRO) framework, in which systems oscillate between marginal and deep contact while simultaneously losing angular momentum through magnetic braking (\citealt{flannery1976cyclic, wang1994thermal, qian2001orbital}).

Contact binaries exhibit continuous light variations and eclipse minima of similar depth, indicating nearly equal surface temperatures (\citealt{1968ApJ...151.1123L}; \citealt{1976ApJ...205..208L}). Their orbital periods typically remain below one day, most often between 0.2 and 0.6 days (\citealt{2006Ap.....49..358D}; \citealt{2021ApJS..254...10L}). Although all-sky surveys reveal that these systems are common (\citealt{2002PASP..114.1124R}), many aspects of their structure and evolution remain unresolved, especially regarding the combined roles of magnetic activity, AML, and mass transfer. Magnetic braking and mass loss are now regarded as essential drivers that regulate the secular evolution of W UMa-type systems.

A notable photometric feature frequently identified in contact binaries is the O'Connell effect, characterized by unequal maxima between eclipses \citep{1951PRCO....2...85O}. Current interpretations associate this asymmetry with magnetic activity cycles, large-scale starspots, and mass-flow irregularities (\citealt{2003ChJAA...3..142L}).

Contact binary systems are commonly classified into A and W subtypes based on the relative masses and effective temperatures of their components. In A-subtype systems, the more massive star is also the hotter one. In W-subtype systems, the more massive component has a lower surface temperature. W-subtype binaries typically show stronger signatures of angular momentum loss, more pronounced magnetic activity, and shallower degrees of contact, all of which influence the efficiency and direction of mass transfer (\citealt{1995MNRAS.274.1019S, 2006MNRAS.370L..29G}). According to \cite{2020MNRAS.492.4112Z}, these differences suggest that A and W subtypes may follow distinct evolutionary pathways (\citealt{2005ApJ...629.1055Y, 2020MNRAS.492.4112Z}).

This study provides our first analysis of the light curve of the EZ Oct binary system and describes its physical characteristics. EZ Oct, with its intermediate mass ratio and short period, represents a crucial system for examining the transition between marginal and deep-contact states predicted by TRO theory. Its configuration makes it particularly valuable for probing the balance between mass transfer and AML-driven evolution in W UMa binaries. Therefore, understanding its photometric behavior and period variations can provide valuable constraints on its evolutionary state and the physical processes governing W UMa-type binaries. Additionally, the noticeable light curve asymmetries of EZ Oct indicate that understanding the possible role of magnetic cycles or starspot activity is essential for interpreting its photometric and orbital behavior.

This work continues the investigation by incorporating new observations and carrying out a detailed analysis of additional W Ursae Majoris (W UMa)-type contact binaries in the BSN\footnote{\url{https://bsnp.info/}} project. The structure of the paper is as follows: Section 2 outlines the selected EZ Oct binary system, the photometric observations, and the data reduction process. Section 3 presents the times of minima and the new ephemeris of EZ Oct. The photometric light curve analysis is presented in Section 4. Section 5 describes the method used to determine the absolute parameters of the system. Finally, Section 6 provides the discussion and conclusions.

\vspace{0.6cm}
\section{System Overview and Observing Process}
The EZ Oct (GSC 09517-00107) binary system has an apparent magnitude of $V^{\mathrm{mag}}=$11.57(11)\footnote{\url{http://simbad.cds.unistra.fr/simbad}} and is located in the southern hemisphere with coordinates R.A.: $15^h$ $43^m$ $13.839386^s$ and Dec: $-86^\circ$ $48'$ $07.227261"$ (J2000). This system is identified as an eclipsing W Ursae Majoris (EW)-type by \cite{2006IBVS.5674....1O}. Although this system is currently classified as a contact binary in various catalogs and databases, including the Variable Star Index (VSX\footnote{\url{https://www.aavso.org/vsx/}}) database with 	0.285878 days orbital period, it was initially identified by the All-Sky Automated Survey for Supernovae (ASAS; \citealt{2018MNRAS.477.3145J}) catalog as a $\delta$ Scuti star with a period of 0.142939 days (\citealt{2006IBVS.5674....1O}). The ASAS catalog has since updated its classification to recognize the system as a contact binary. The VSX and ASAS-SN databases report an orbital period of 0.285878 days and maximum apparent magnitudes of $11.37^{\mathrm{mag}}$ and $11.56^{\mathrm{mag}}$ in the $V$ band, respectively.
\\
\\
EZ Oct was observed on May 25, 2023, August 9, 2024, and July 22, 2025, using the 2.15-meter Jorge Sahade Telescope (JS) at the Complejo Astronómico El Leoncito Observatory (CASLEO), located in San Juan, Argentina ($69^{\circ}18^{\prime}$W, $31^{\circ}48^{\prime}$S; 2552~m above sea level). A total of 926 images were acquired during the observing process. The observations were carried out using a VersArray 2048B CCD camera (Roper Scientific, Princeton Instruments), operated in a 5~$\times$~5 pixel binning mode to enhance the signal-to-noise ratio and reduce readout time. Standard Johnson-Cousins $B$, $V$, $R_c$, and $I_c$ filters were used throughout the observations, with an exposure time of 12 seconds per frame.

The data reduction and aperture photometry were performed using the APPHOT package within the Image Reduction and Analysis Facility (IRAF; \citealt{1986SPIE..627..733T}). The reduction process incorporated the use of bias and flat-field calibration frames to ensure accurate photometric measurements.
\\
\\
High-quality timeseries photometric data from the Transiting Exoplanet Survey Satellite (TESS) was available for EZ Oct (TIC 290410988). The TESS mission primarily aims to identify and categorize exoplanets through systematic observations and data analysis (\citealt{2015JATIS...1a4003R}). Each TESS sector monitors a designated portion of the sky for approximately 27.4 days. All observational data used in this study were retrieved from the Mikulski Archive for Space Telescopes (MAST)\footnote{\url{https://mast.stsci.edu/portal/Mashup/Clients/Mast/Portal.htmL}}. TESS-like light curves were generated using the LightKurve package\footnote{\url{https://docs.lightkurve.org}}, and the data were processed to remove trends by applying the TESS Science Processing Operations Center (SPOC) pipeline (\citealt{2016SPIE.9913E..3EJ}).

Sector 66 was selected for light curve analysis as it contains the most recent available time-series data. To calculate a new ephemeris, observations from all available TESS sectors were incorporated. The TESS sectors used, along with their exposure times and average flux uncertainties, are listed in Table \ref{tess}.

\begin{table}
\caption{Details of the TESS dataset adopted in this research.}
\centering
\begin{center}
\footnotesize
\begin{tabular}{c c c c c}
\hline
TESS & Observation & Exposure & Error\\
Sector & Year & Length & Average\\
\hline
12 & 2019 & 120 & 0.0013\\
13 & 2019 & 120 & 0.0014\\
65 & 2023 & 200 & 0.0012\\
66 & 2023 & 200 & 0.0011\\
\hline
\end{tabular}
\end{center}
\label{tess}
\end{table}

\vspace{0.6cm}
\section{Orbital Period Variation}
This work provides a comprehensive analysis of EZ Oct using updated and available data. We compiled as many times of minima as possible from photometric surveys to support the analysis of orbital period variations in the system. Since the dataset included both Barycentric Julian Dates in Barycentric Dynamical Time ($BJD_{TDB}$) and Heliocentric Julian Dates ($HJD$), we first standardized all entries to $BJD_{TDB}$ using an online conversion tool\footnote{\url{https://astroutils.astronomy.osu.edu/time/hjd2bjd.html}}. We determined the primary and secondary times of minima from our observations in the employed filters, as well as from the available TESS timeseries data. The times of minima derived from our observations are listed in Table \ref{min}, while a machine-readable version of the minima extracted from the TESS data is also provided online.

To calculate the epoch and observed-minus-calculated (O-C) values for each eclipse time, we used the orbital period of 0.285878 days from the VSX database, adopting the minimum time from our observation as the reference ephemeris (Table \ref{min} and a machine-readable version Table).

\begin{figure*}
\centering
\includegraphics[width=\linewidth]{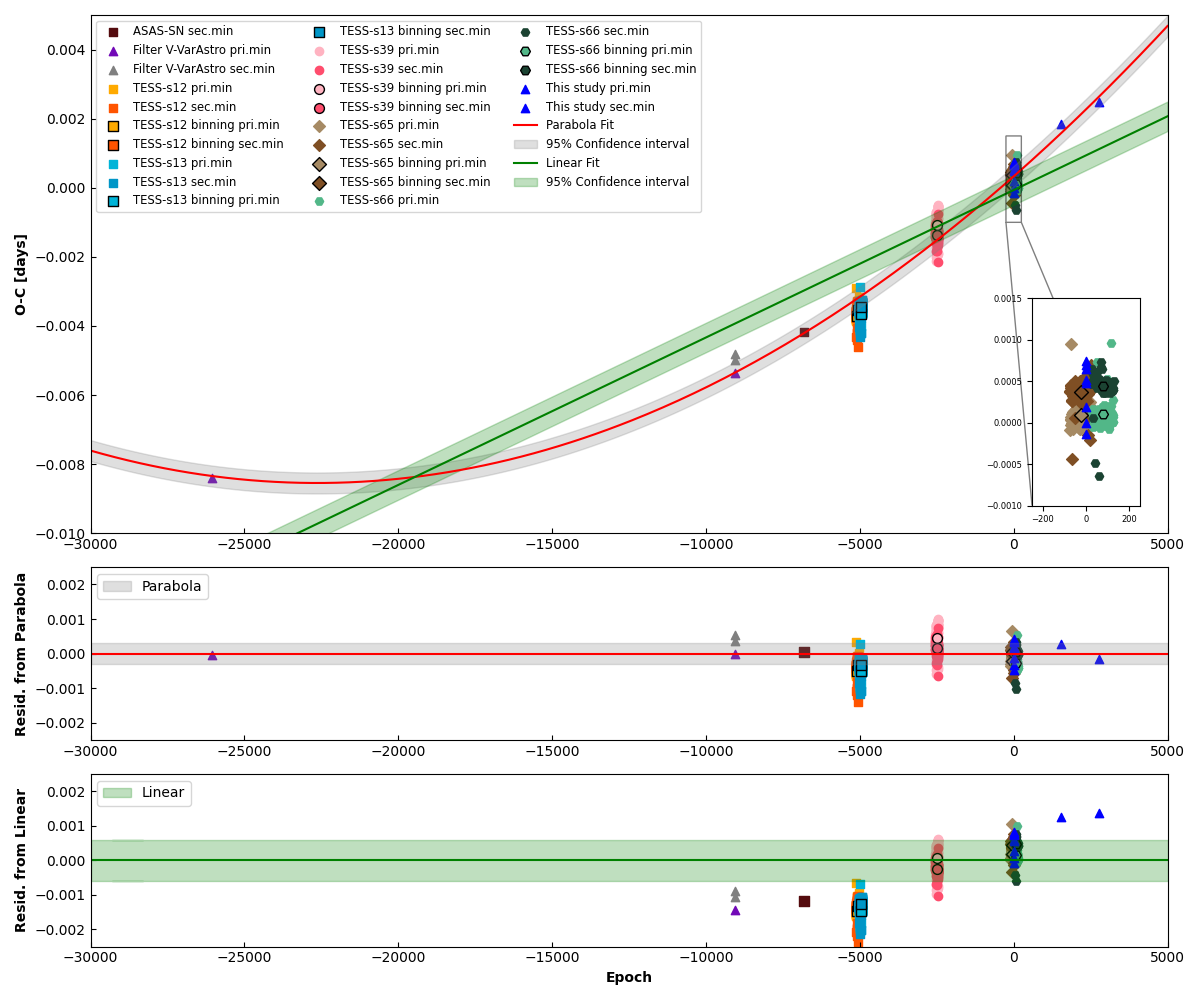}
\caption{O–C variations of EZ Oct. The diagram is based on a total of 946 minima, including those collected from VarAstro and ASAS-SN, as well as minima extracted from various TESS sectors and our ground-based observations. The linear and parabolic fits are shown on the data, and in the lower panels, their corresponding residuals are displayed.}
\label{OC}
\end{figure*}

The O-C diagram is a fundamental tool for diagnosing orbital period variations in eclipsing binary systems. For the EZ Oct system, we constructed a detailed O-C diagram (Figure \ref{OC}) based on eclipse timing data collected over several years. Figure \ref{OC} also includes an bottom panel that highlights the residuals obtained after subtracting the quadratic fit. This extensive dataset incorporates high-precision measurements from both space-based missions and complementary ground-based photometric observations. The resulting temporal baseline enables a sensitive probe of small-scale orbital variations, providing critical insight into the system's dynamical evolution.

In order to determine which model provides the most adequate description of the O–C variations of EZ Oct and should be adopted as the final O–C trend, we compared the linear and quadratic fits using the Akaike Information Criterion (AIC) and the Bayesian Information Criterion (BIC). For the quadratic fit, the values AIC = 367 and BIC = 368 were obtained, while the linear fit resulted in AIC = 14513 and BIC = 14503. Since smaller values of both criteria indicate a more suitable model, these results clearly favor the quadratic fit. Moreover, as evident from the O–C diagram (Figure \ref{OC}), the majority of the data points are concentrated between epochs –9044 and 128, with only a single data point located at epoch –26049, corresponding to the observations reported by VarAstro\footnote{\url{http://var2.astro.cz/ocgate/}}. Even after excluding this point from the analysis, the superiority of the quadratic model remains evident. The quadratic fit yields AIC = 4865 and BIC = 4879, whereas the linear fit gives AIC = -15285 and BIC = -15275, indicating a substantially better statistical performance for the quadratic solution. This pronounced difference between the information criteria confirms the robustness of the quadratic O-C trend. Therefore, the quadratic model was adopted as the final O-C variation for EZ Oct.

Analysis of the O-C diagram for EZ Oct revealed a pronounced upward curvature, indicative of a sustained secular increase in its orbital period. To quantify this behavior, a quadratic least-squares fit was applied to the eclipse timing data using the ephemeris equation. Hence, the quadratic ephemeris of EZ Oct is expressed as:

\begin{equation}
T_{min}=T_0+P_0E+aE^2
\end{equation}

\noindent where $T_0$ and $P_0$ come from the reference epoch, and $a$ represents the quadratic period variation coefficient, with a value of $a = 1.7268 \times 10^{-11}$.

We derived light elements from the minima measured in this study, together with TESS data and values from the literature, and used them to establish an updated ephemeris:

\begin{equation}
\label{eq1}\begin{aligned}
BJD_{TDB}(Min.I)=2460089.60443(25)+0.28587864(63)\times E
\end{aligned}
\end{equation}

\noindent where $E$ denotes the number of orbital cycles elapsed since the reference epoch.

Moreover, the instantaneous period change rate is defined by:

\begin{equation}
\dot{P}  \approx \frac{dP}{dE} . \frac{dE}{dt}= 2aP_0
\end{equation}

Substituting the best-fit coefficients yields a value of \(\dot{P}  \approx 1.090\times 10^{-3}\) sec/year.

This statistically significant positive period derivative is consistent with theoretical expectations for conservative transfer of mass and energy from the less massive star to its more massive companion, resulting in an expanding orbital separation and increasing orbital period (\citealt{2005ApJ...629.1055Y}).

\begin{table*}
\caption{Collected and extracted ground-based times of minima.}
\centering
\begin{center}
\footnotesize
\begin{tabular}{c c c c c c}
\hline
Min.($BJD_{TDB}$) & Error & Epoch & O-C(day) & Filter & Reference\\
\hline
2452642.75976 & & -26049 & -0.00839 & $V$ & VarAstro\\
2457504.12354	&	&	-9044	&	-0.00535	& $V$ & VarAstro\\
2457504.26648	&	&	-9043.5	&	-0.00499	& $V$ & VarAstro\\
2458138.05800	&	&	-6826.5	&	-0.00416	& $V$ & ASAS-SN	\\
2460089.60403	&	0.00015	&	0.0	&	-0.00014	& $R$ & This study	\\
2460089.60417	&	0.00019	&	0.0	&	0.00000	&	 $V$ & This study	\\
2460089.60436	&	0.00028	&	0.0	&	0.00019	& $I$ & This study	\\
2460089.60491	&	0.00022	&	0.0	&	0.00074	& $B$ & This study	\\
2460089.74759	&	0.00011	&	0.5	&	0.00048	&	 $V$ & This study	\\
2460089.74762	&	0.00008	&	0.5	&	0.00051	&	 $I$ & This study	\\
2460089.74776	&	0.00007	&	0.5	&	0.00065	&	 $B$ & This study	\\
2460089.74780	&	0.00011	&	0.5	&	0.00069	&	 $R$ & This study	\\
2460531.57341	& 0.00190	& 1546	& 0.00185 & $V$ & This study	\\
2460878.62992	& 0.00170	& 2760	& 0.00247 & $V$ & This study	\\
\hline
\end{tabular}
\end{center}
\label{min}
\end{table*}

\vspace{0.6cm}
\section{Light Curve Solution}
To analyze the photometric light curves of the EZ Oct binary system, we used version 2.4.9 of the PHysics Of Eclipsing BinariEs (PHOEBE) Python package \citep{2016ApJS..227...29P, 2020ApJS..250...34C}. The observation times in the light curves were converted into orbital phases based on the reference ephemeris provided in Section 3. Given the morphology of the observed light curves, the systems' classifications in the catalogs, and their relatively short orbital periods, the contact binary configuration was selected for the modeling process. The gravity darkening parameter was fixed at $g_1 = g_2 = 0.32$ (\citealt{1967ZA.....65...89L}), while the bolometric albedo was set to $A_1 = A_2 = 0.5$ (\citealt{1969AcA....19..245R}). Stellar atmospheres were modeled using the approach described by \cite{2004AA...419..725C}, and the limb darkening coefficients were left as adjustable parameters within PHOEBE.

The initial effective temperature ($T$) used in the analysis was obtained from the Gaia DR3 catalog, which reports a value of 5025 K for this system. This temperature corresponds to the hotter component, as determined from the depth of the light curve minima. The initial effective temperature of the cooler star was inferred from the difference in depth between the primary and secondary minima of the light curves.

\begin{figure}
    \centering
    \includegraphics[scale=0.09]{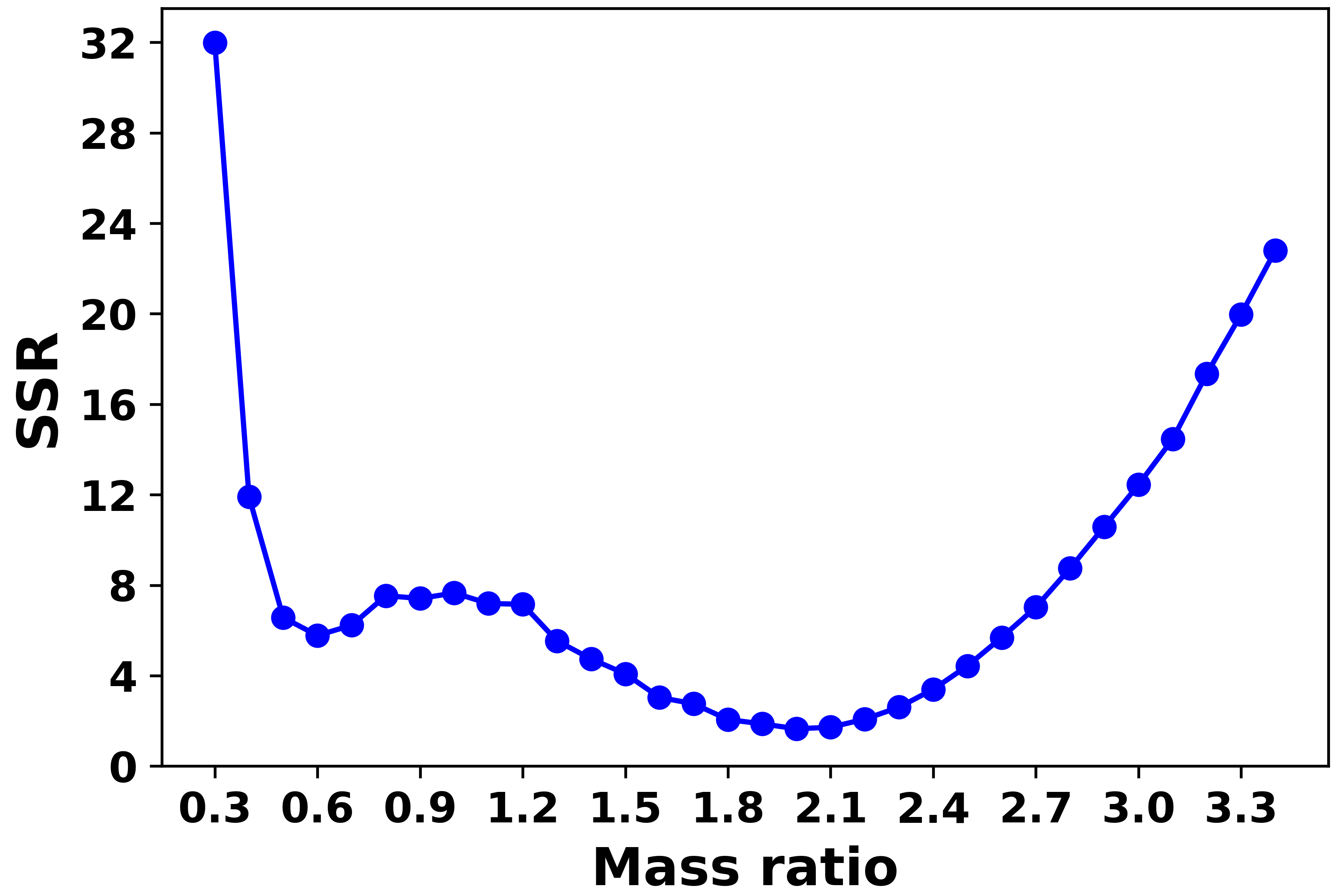}
    \caption{Sum of the squared residuals as a function of the mass ratio.}
    \label{q}
\end{figure}

The initial mass ratio of the EZ Oct binary system was derived through the $q$-search approach \citep{2005ApSS.296..221T}. A wide span of mass ratio values, ranging from 0.1 to 20, was explored during the $q$-search. This was followed by a more focused search within a narrower interval to refine the result by minimizing the sum of squared residuals. As shown in Figure \ref{q}, the $q$-search curve display a distinct minimum, indicating the most likely solution. Previous studies, such as those by \cite{2021AJ....162...13L} and \cite{2024AJ....168..272P}, have demonstrated that systems with high orbital inclinations and total eclipses yield reliable mass ratio determinations.

\begin{figure}
\centering
\includegraphics[scale=0.2]{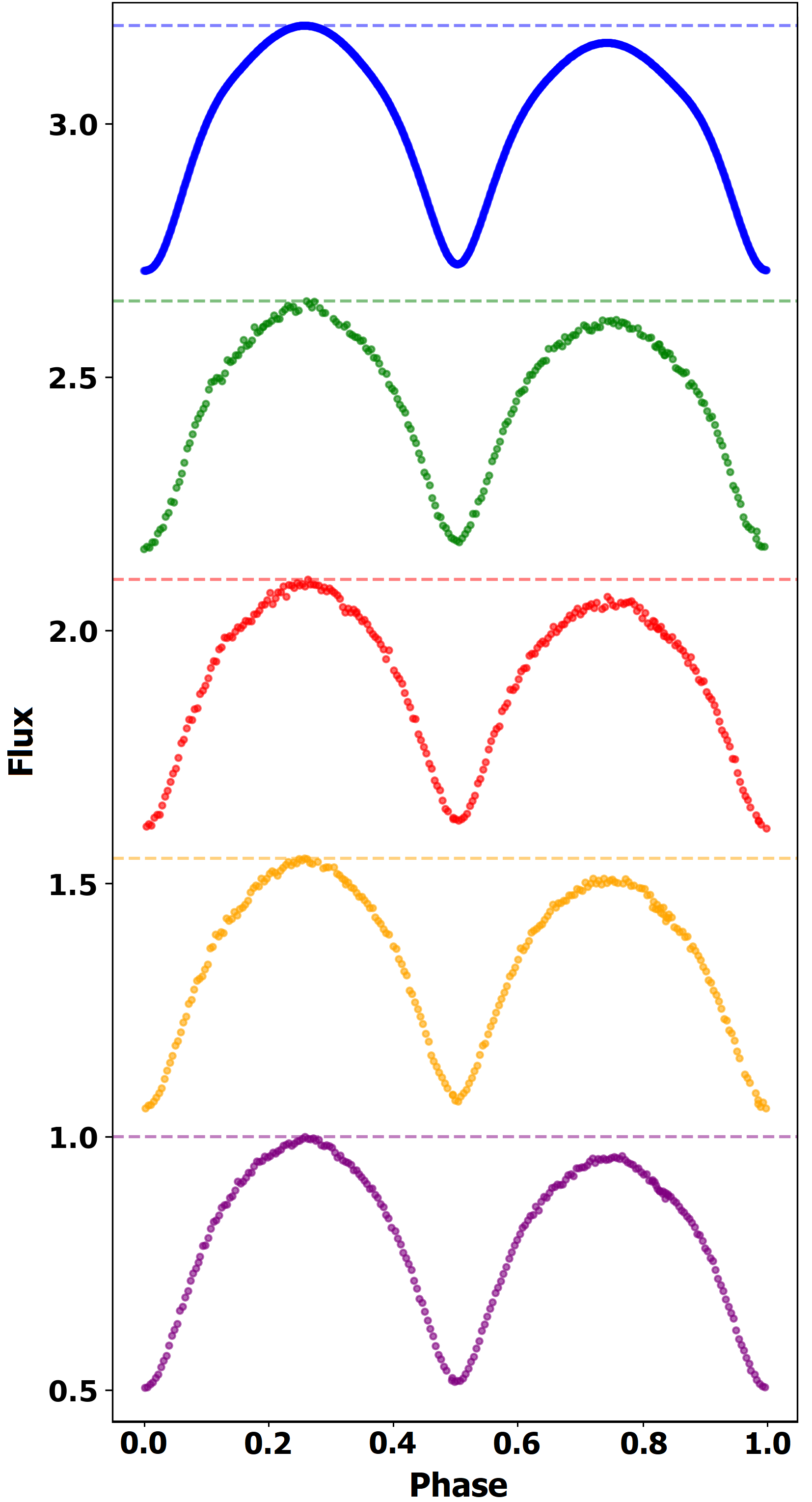}
\caption{Phase–Flux observational light curves for five filters: $B$ (blue), $V$ (green), $R$ (red), $I$ (orange), and TESS (purple), arranged from top to bottom with offsets applied for clarity. Horizontal dashed lines indicate the higher of the two maxima for each light curve, highlighting unequal maxima.}
\label{lc-O'Connell effect}
\end{figure}

The analysis reveals that EZ Oct displays unequal maxima in its light curve. The flux differences between the two maxima are 0.037, 0.035, 0.036, 0.032, and 0.034 in the $B$, $V$, $R$, $I$, and TESS filters, respectively (Figure \ref{lc-O'Connell effect}). To account for this asymmetry in the light curve solution process, a cold starspot was required on one of the stellar components. This feature is most likely the result of magnetic activity, causing regions of higher or lower temperature on the stellar surface and manifesting as the well-known O'Connell effect (\citealt{1951PRCO....2...85O},  \citealt{2017AJ....153..231S}). Including a starspot to account for the O'Connell effect improves the fit, with a reduced $\chi^2 \approx 5.12$, compared to $\chi^2 \approx 14.38$ without the starspot. Table \ref{analysis} lists the starspot properties: colatitude, longitude, angular radius, and temperature ratio.

The evolution of the spot across the four available TESS sectors provides a clear explanation for the long-term variation observed in the O'Connell effect. The difference between the two maxima was measured individually for each sector. In the 2019 data, the asymmetry is relatively weak, with $\Delta{\rm Max} = 0.0137$ in Sector 12 and $\Delta{\rm Max} = 0.0219$ in Sector 13. In contrast, the 2023 observations show a significantly stronger asymmetry, with $\Delta{\rm Max} = 0.0276$ in Sector 65 and $\Delta{\rm Max} = 0.0381$ in Sector 66. This clear increase demonstrates that the light curve asymmetry becomes substantially stronger over the four-year interval. Each sector was modeled independently to investigate this behavior. The photometric solutions show that the physical and geometric parameters of the system remain unchanged across all epochs, and that the observed variations originate solely from the evolution of the spot configuration. The derived spot parameters for all sectors are summarized in Table \ref{tab:spot_evolution}. The modeling reveals a consistent and physically meaningful temporal evolution of the spot properties. In all sectors, the spot is located on the secondary component at a nearly constant latitude of approximately $71^\circ$, suggesting stable magnetic activity within the same latitude band. The spot longitude exhibits a gradual forward migration, increasing from $260^\circ$ in Sector~12 to $269^\circ$ in Sector~66, corresponding to a drift of about $9^\circ$ over four years. In addition, the spot radius increases steadily from $18^\circ$ in Sector~12 to $24^\circ$ in Sector~66, while the temperature factor remains constant at $T_{\rm spot}/T_{\rm surf} = 0.89$, indicating that the thermal contrast of the spot does not change. This combination of longitudinal migration, progressive enlargement of the spot, and constant thermal contrast naturally reproduces the observed increase in the O’Connell effect. The transition from the relatively weak asymmetry in 2019 to the much stronger asymmetry in 2023 can therefore be fully explained by geometric and areal evolution of a cool spot on the secondary star, without requiring any modification to the structural or orbital parameters of the binary. This behavior is consistent with long-term magnetic activity cycles commonly observed in contact binaries.

The observed light curve was modeled using photometric multiband data and initial parameter estimates to achieve a satisfactory theoretical fit. The optimization tool in PHOEBE was subsequently applied to further refine the solution and improve the accuracy of the light curve modeling.

To determine the final parameter values and their associated uncertainties, we applied the Markov Chain Monte Carlo (MCMC) method using the emcee package (\citealt{2013PASP..125..306F}). While the BSN application (\citealt{paki2025bsn}) was employed to derive an initial synthetic fit prior to conducting the MCMC modeling. The five key parameters included in the MCMC modeling were the mass ratio ($q$), orbital inclination ($i$), fillout factor ($f$), and the effective temperatures of both stellar components ($T_1$ and $T_2$). A Gaussian likelihood function was adopted to represent the observational light curve comprehensively. The sampling was carried out using 32 walkers over 1000 iterations, with the initial 300 steps discarded as burn-in. The results of the light curve analysis showed no evidence of a third light contribution ($l_3$) in any of the systems studied.

\begin{figure*}
    \centering
    \includegraphics[scale=0.44]{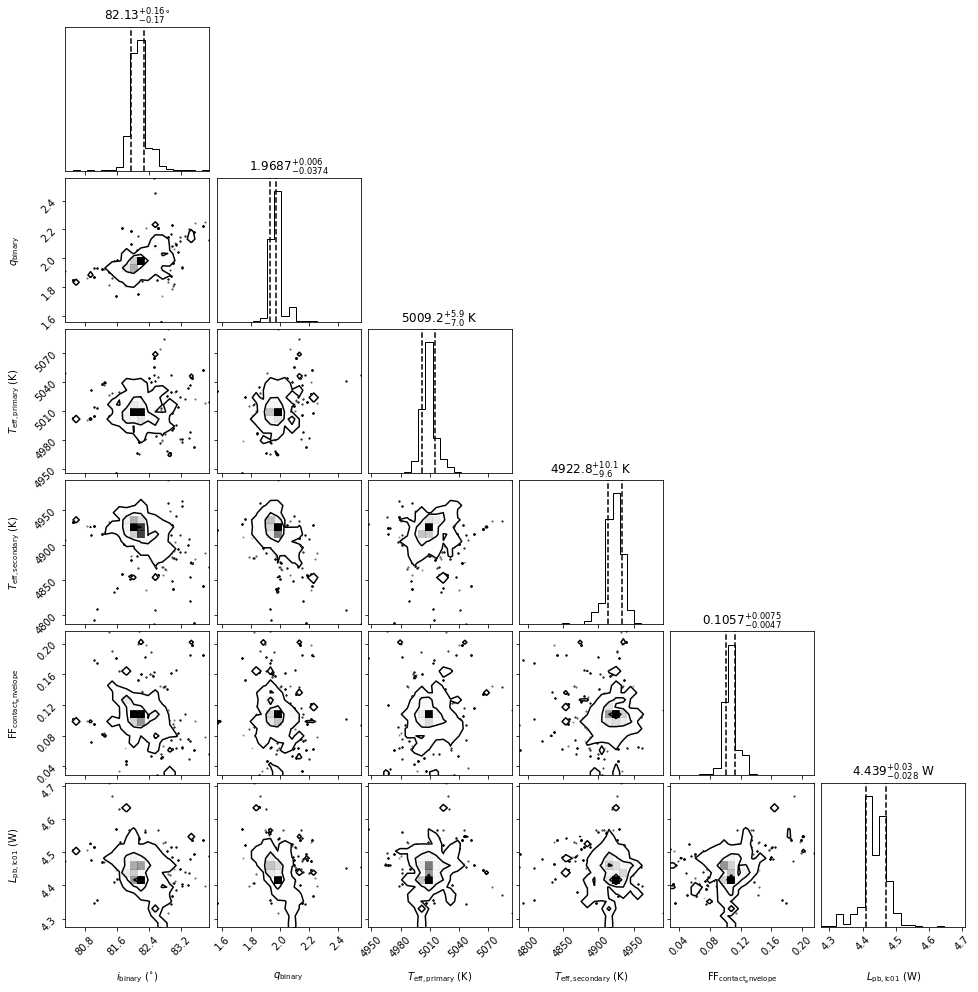}
    \caption{MCMC-derived corner plots for the EZ Oct system.}
    \label{corner}
\end{figure*}

\begin{figure*}
    \centering
    \includegraphics[width=\linewidth]{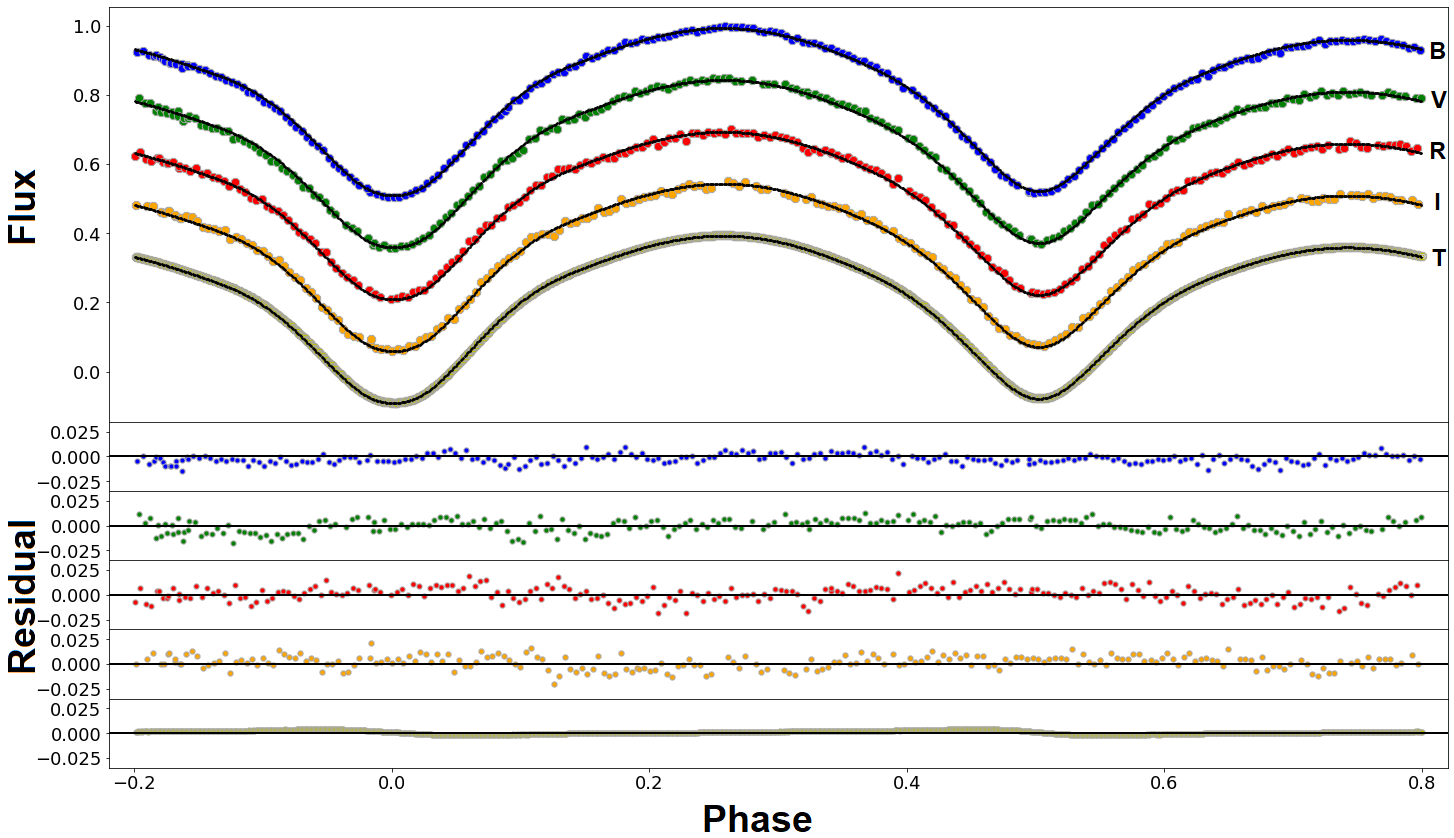}
    \caption{Observed (colored dots) and modeled (solid lines) light curves for the systems across $BVR_cI_c$ and TESS filters, arranged from top to bottom.}
    \label{LC}
\end{figure*}

\begin{figure*}
\centering
\includegraphics[width=\linewidth]{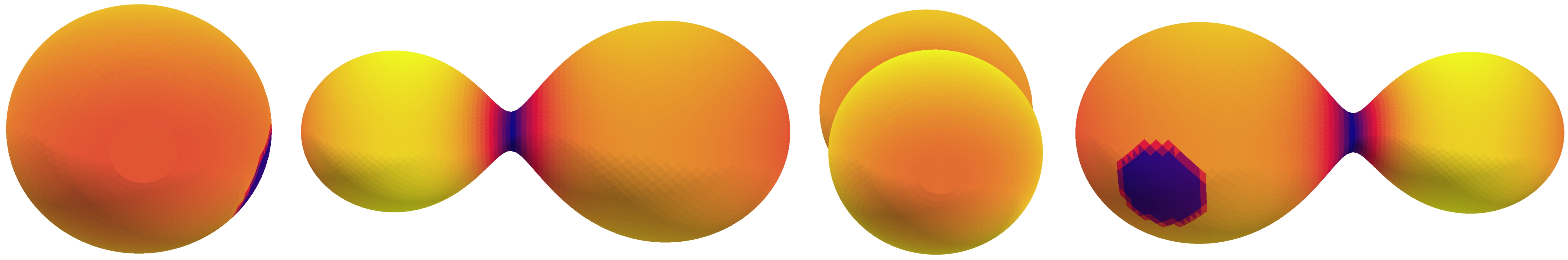}
\includegraphics[width=\linewidth]{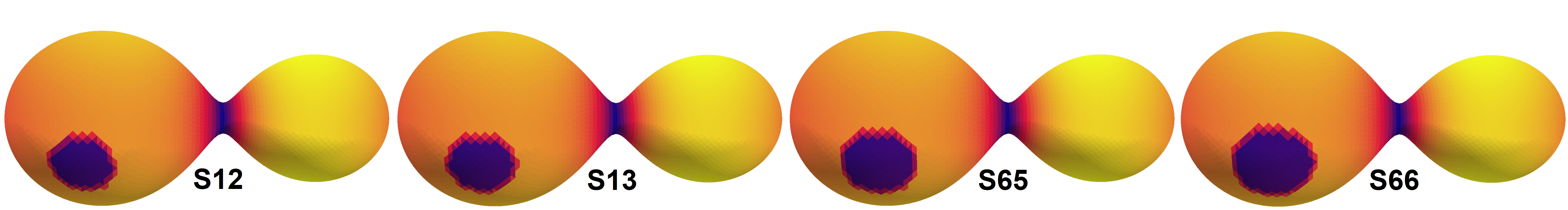}
\caption{Top: Views of the EZ Oct binary system in 3D at four orbital phases 0, 0.25, 0.50, and 0.75, respectively. Bottom: 3D view of both components at orbital phase 0.75 for the four TESS sectors, from left to right. The apparent variations of the cool stellar spot on the secondary star over time are clearly visible.}
\label{3D}
\end{figure*}

The results of the light curve analysis for EZ Oct are summarized in Table \ref{analysis}. The corresponding corner plot is shown in Figure \ref{corner}. Figure \ref{LC} presents the final synthetic light curves overlaid with the observed data for the binary systems. Additionally, three-dimensional representations of the systems are illustrated in Figure \ref{3D}.

\begin{table*}
\renewcommand\arraystretch{1.5}
\caption{Estimated photometric parameters of EZ Oct derived from the light curve analysis.}
\centering
\begin{center}
\footnotesize
\begin{tabular}{c c c c c c c}
\hline
Parameter && Result && Parameter && Result\\
\hline
$T_{1}$ (K) && $5009_{\rm-7}^{+6}$ &&  $r_{1(mean)}$ && ${0.330 \pm 0.008}$ \\
$T_{2}$ (K) && $4923_{\rm-10}^{+10}$ && $r_{2(mean)}$ && ${0.448 \pm 0.007}$\\
$q=M_2/M_1$ && $1.969_{\rm-0.037}^{+0.006}$ && Phase shift && $0.024$ \\
$f$ && $0.106_{\rm-0.005}^{+0.008}$ && Colatitude$_{spot}$(deg) && $71$\\
$i^{\circ}$ &&	$82.13_{\rm-0.17}^{+0.16}$ && Longitude$_{spot}$(deg) && $256$\\
$\Omega_1=\Omega_2$ && ${5.145 \pm 0.063}$ && Radius$_{spot}$(deg) && $21$ \\
$l_1/l_{tot}$ && $0.375_{\rm-0.002}^{+0.002}$  && $T_{spot}/T_{star}$ && $0.89$ \\
$l_2/l_{tot}$ && ${0.625 \pm 0.002}$ && Component$_{spot}$ && Secondary \\
\hline
\end{tabular}
\end{center}
\label{analysis}
\end{table*}

\begin{table}
\centering
\caption{Starspot parameters obtained from modeling the four TESS sectors.}
\label{tab:spot_evolution}
\begin{tabular}{ccccc}
\hline
TESS Sector & Latitude$^\circ$ & Longitude$^\circ$ & Radius$^\circ$ & $T_{\rm spot}/T_{\rm surf}$ \\
\hline
12 & 71 & 250 & 18 & 0.89 \\
13 & 71 & 254 & 19 & 0.89 \\
65 & 71 & 257 & 21 & 0.89 \\
66 & 71 & 259 & 23 & 0.89 \\
\hline
\end{tabular}
\end{table}

\vspace{0.6cm}
\section{Fundamental Parameters}
We estimated the absolute parameters of the EZ Oct system using the parallax reported in Gaia DR3 (\citealt{2024NewA..11002227P}). This approach is considered reliable when photometric observations are available, due to the high precision of Gaia DR3 parallaxes (\citealt{2021AJ....162...13L}). However, to ensure the validity of this method, certain conditions must be met: the visual extinction coefficient ($A_V$) should be less than approximately 0.4 (\citealt{2024PASP..136b4201P}), and the Renormalized Unit Weight Error (RUWE) must not exceed 1.4 (\citealt{lindegren2018re}). For the EZ Oct system, both criteria are satisfied. We determined $A_V=0.220\pm 0.001$ using the 3D dust map by \cite{2019ApJ...887...93G}, and the RUWE value was taken directly from the Gaia DR3 archive. Additionally, we adopted $V_\mathrm{max}=11.57\pm 0.04$ comes from our observations as the system's apparent brightness at maximum light.

First, the system's absolute visual magnitude ($M_V$) was calculated from its apparent magnitude ($V$), the Gaia DR3-based distance, and the visual extinction ($A_V$). Subsequently, the individual absolute magnitudes of the components, $M_{V1}$ and $M_{V2}$, were determined using the light ratio in the $V$-band ($l_{1,2}/l_{\mathrm{tot}}$) obtained from the light curve solution. The absolute bolometric magnitudes ($M_{\mathrm{bol1,2}}$) were then calculated by applying bolometric corrections ($BC_{1,2}$) taken from the study by \citealt{1996ApJ...469..355F}. The radii of the binary components were derived from the bolometric magnitudes and the corresponding luminosities ($L$). In this calculation, we adopted a solar bolometric magnitude of $M_{\mathrm{bol}\odot} = 4.73$ mag, following \citealt{2010AJ....140.1158T}. With the luminosity estimated in this way and the effective temperatures obtained from the light curve analysis, the stellar radii ($R$) were computed. The semi-major axis $a (R_{\odot})$ was determined using the radii $R_{1,2}$ and the mean relative radii $r_{\mathrm{mean1,2}}$, by averaging the resulting values of $a_1(R_{\odot})$ and $a_2(R_{\odot})$. Finally, the individual stellar masses were calculated from $a$, $P_{\mathrm{orb}}$, and $q$ using Kepler's third law (Equations \ref{eq:M1}, \ref{eq:M2}).

\begin{eqnarray}
M{_1}=\frac{4\pi^2a^3}{GP^2(1+q)}\label{eq:M1}\\
M{_2}=q\times{M{_1}}\label{eq:M2}
\end{eqnarray}

The derived absolute parameters for EZ Oct are presented in Table \ref{fundamental}.

\begin{table*}
\renewcommand\arraystretch{1.2}
\caption {Estimation of absolute parameters.}
\centering
\begin{center}
\footnotesize
\begin{tabular}{c c c}
\hline
Parameter & Hotter star & Cooler star\\
\hline
$M(M_\odot)$& ${0.454 \pm 0.006}$&  $0.895\pm0.011$\\ 
$R(R_\odot)$& ${0.665 \pm 0.015}$& $0.906\pm0.024$\\
$L(L_\odot)$& ${0.250 \pm 0.010}$& $0.433\pm0.019$\\
$M_{bol}(\text mag)$& ${6.245 \pm 0.043}$&  $5.65\pm0.048$\\
$log(g)(\text cgs)$& ${4.449 \pm 0.025}$&  $4.476\pm0.028$\\
$a(R_\odot)$  &  \multicolumn{2}{c}{$2.018 \pm0.008$}\\
\hline
$BC(\text mag)$ & $-0.304 \pm 0.003$ & $-0.344 \pm 0.005$\\
\hline
\end{tabular}
\end{center}
\label{fundamental}
\end{table*}

\vspace{0.6cm}
\section{Discussion and Conclusion}
This study provides a detailed photometric analysis of the contact binary system EZ Oct, encompassing orbital period variation studies, light curve modeling, and determination of absolute stellar parameters. The data were obtained through multiband photometric observations conducted at an observatory in Argentina. Based on the results, the following discussion and conclusions are presented:

A) A defining characteristic of overcontact binary systems such as EZ Oct is the sustained and dynamically significant mass exchange between their stellar components. This process is a key driver of orbital evolution, influencing both the orbital period and the system's angular momentum budget, and ultimately steering the binary toward a coalescence phase.

Assuming conservative mass transfer—in which no mass escapes the system—the rate of orbital period variation is directly linked to the mass transfer rate via the following relation (\citealt{2001icbs.book.....H}):

\begin{equation}
\begin{aligned}
\frac{\dot{P}}{P}=-3\dot{M}(\frac{1}{M_1}-\frac{1}{M_2}).
\end{aligned}
\end{equation}

Using this relation, we estimate the mass transfer rate as $\dot{M}=1.353\times 10^{-8}$ $M_{\odot}$/year.

This substantial rate indicates rapid mass redistribution from the less massive primary to the more massive secondary. Under conservative conditions, such mass flow increases the orbital separation and period, consistent with the upward curvature observed in the O–C diagram (Figure \ref{OC}).

It is worth noting that no significant cyclic modulation was detected in the O–C residuals beyond the fitted parabolic trend. This suggests that if a third body is present, it either has a long orbital period beyond the current observational baseline or exerts negligible dynamical influence on the inner binary. Additionally, there is no strong evidence for short-term variations due to magnetic activity cycles (e.g., the Applegate mechanism, \citealt{1992ApJ...385..621A}), though such effects may become apparent with higher-cadence, long-term observations in the future.

Furthermore, the Thermal Relaxation Oscillation (TRO) model, as described in recent studies (e.g., \citealt{2025AJ....169...85X}), provides a theoretical framework for cyclic changes in overcontact binaries. TRO predicts intermediate-term oscillations in the envelope structure, leading to periodic variations in the orbital period. Similar analyses in comparable systems indicate that these cycles can produce observable modulations in the O-C diagram. For EZ Oct, despite having 946 times of minima from various sources, the temporal coverage is still limited, and light curve parameters such as fillout factor, mass ratio, and temperature differences have been derived only for restricted epochs (\citealt{2024NewA..10502112S}). Consequently, while the quadratic O-C trend can be fitted, the current data do not provide sufficient temporal baseline or resolution to robustly identify or quantify TRO cycles. Nevertheless, the observed upward curvature in the O-C diagram is broadly consistent with expectations from TRO-induced period variations. Long-term, high-cadence monitoring will be essential to explore these effects quantitatively in future studies.

B) In the contact binary system EZ Oct, the observed asymmetry between the two light curve maxima, which is a signature of the O'Connell effect (\citealt{1951PRCO....2...85O}), required the addition of a cool starspot on one of the stellar components in order to achieve a satisfactory fit.
Moreover, the light curve solution indicates a temperature difference of 86 K between the two stellar components. The final effective temperature of EZ Oct, derived using an initial value from Gaia DR3, is consistent with the overall correlation between Gaia-based temperatures and model-derived values as investigated by \cite{2025MNRAS.537.3160P}.
The spectral classifications of the components were derived using temperature-based criteria from \citet{2000asqu.book.....C} and \citet{2018MNRAS.479.5491E}. Considering their effective temperatures, the components of EZ Oct were classified as K1 and K2 spectral types.

C) In contact binary systems, the extent of envelope sharing between the stellar components is described by the fillout factor ($f$), which quantifies how much the stars overfill their Roche lobes and merge into a common envelope. This parameter serves as a key indicator of the degree of contact, offering insight into the efficiency of mass and energy exchange within the system. Binaries with small $f$ values are typically in a shallow contact configuration, where the shared envelope is limited, while larger values reflect deeper levels of interaction and more significant overcontact. The fillout factor is calculated using the following expression:

\begin{equation}
f = \frac{\Omega - \Omega_{\mathrm{in}}}{\Omega_{\mathrm{out}} - \Omega_{\mathrm{in}}},
\end{equation}

\noindent where $\Omega$ denotes the effective surface potential of the system, and $\Omega_{\mathrm{in}}$ and $\Omega_{\mathrm{out}}$ represent the inner and outer critical potentials, respectively, as defined by \cite{mochnacki1981}.

\begin{figure}
\centering
\includegraphics[scale=0.1]{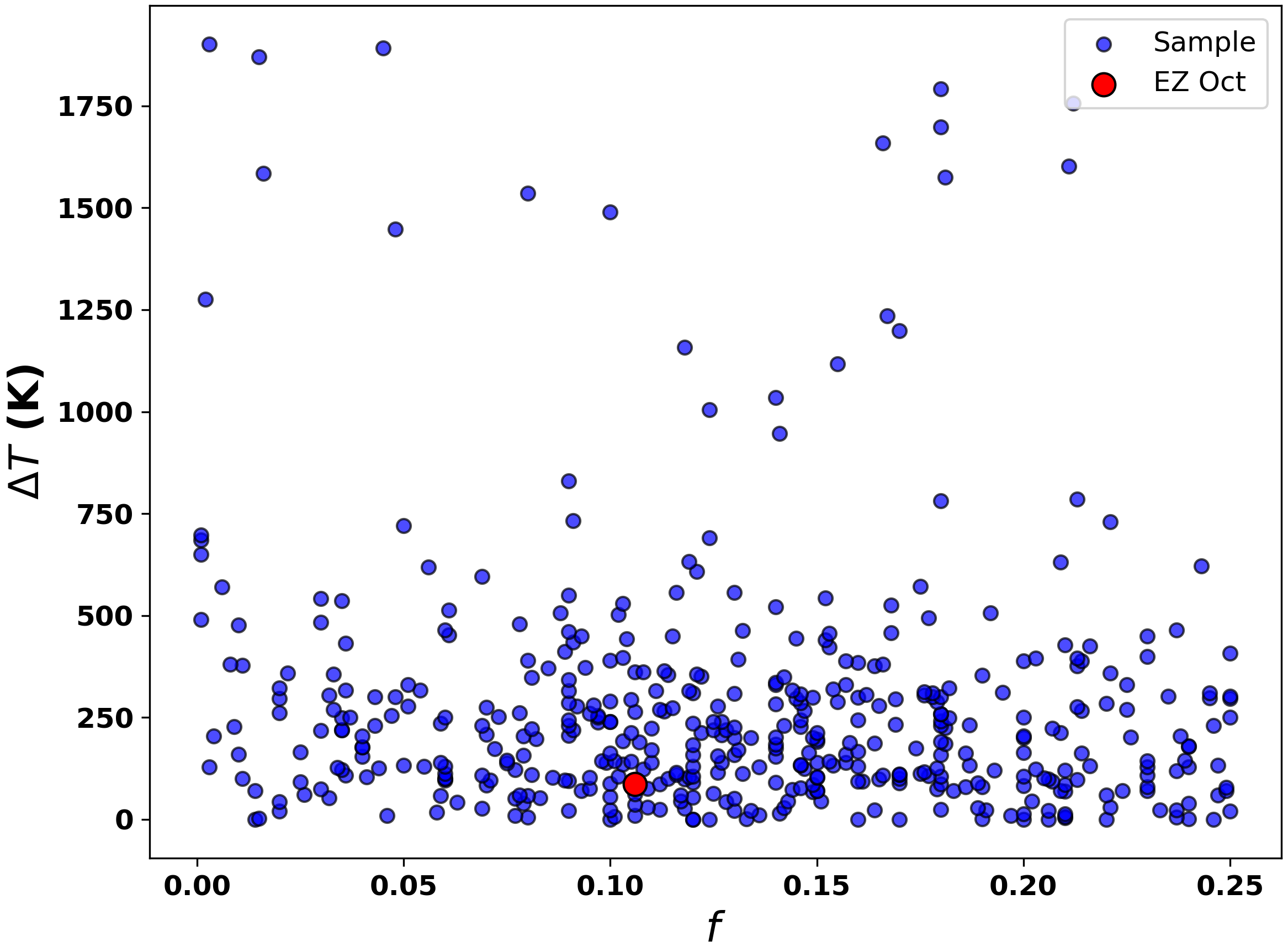}
\caption{Scatter plot of fillout factor ($f$) versus temperature difference ($\Delta T$) for 458 contact binaries with $f < 0.25$. The EZ Oct system is highlighted as a red point.}
\label{fig:f-dT}
\end{figure}

Based on the value of the fillout factor ($f$), contact binary systems are typically classified into three categories (\citealt{2022AJ....164..202L}): shallow ($f < 25\%$), medium ($25\% \leq f < 50\%$), and deep contact systems ($f \geq 50\%$). According to the light curve solution, EZ Oct falls into the shallow contact category. This configuration may indicate that EZ Oct is in an early evolutionary stage of contact, where the stars are gradually adjusting thermally and structurally before potentially reaching deeper overcontact. The shallow contact observed in target was accompanied by a temperature difference of 86 K between the components. In contact binary systems, a temperature difference is not determined solely by the filling factor; it also depends on the nature of the stars prior to contact and their initial temperatures before mass and energy exchange.
To investigate this further, we extracted a sample of 458 contact binary systems with fillout factors $f < 0.25$ from the \cite{2025MNRAS.538.1427P} study. As shown in Figure \ref{fig:f-dT}, no strong correlation is observed between the fillout factor and the temperature difference of the components in these systems. Figure \ref{fig:f-dT} also illustrates that, according to the literature, there are systems with low $f$ values yet relatively large temperature differences of up to 1900 K, while others exhibit virtually no temperature difference. Therefore, similar systems reported in the literature support the low surface temperature difference between the components and the shallow fillout factor observed in EZ Oct.

D) Since all required conditions were met, the system's absolute parameters were derived from the Gaia DR3 parallax. In this approach, the paths of the primary and secondary stars are treated independently, yielding $a_1(R_\odot) = 2.015$ and $a_2(R_\odot) = 2.022$ as their respective semi-major axes. The computed values show good agreement, with a difference of only $\Delta a = |\!a_2 - a_1\!| = 0.007$, which supports the validity of the method. This $\Delta a$ value further confirms the reliability of the light curve solution and the adopted model parameters \citep{2024NewA..11002227P, 2025MNRAS.538.1427P}.

E) The orbital angular momentum ($J_0$) is a fundamental parameter that characterizes the rotational and orbital dynamics of binary systems, providing insight into their stability and evolutionary state. For EZ Oct, based on the computed mass ratio and component masses, it allows investigation of how the system's orbital angular momentum is distributed relative to other contact binaries. The orbital angular momentum of the EZ Oct system was computed using Equation \ref{eqJ0}, as described in the study by \cite{2006MNRAS.373.1483E}, based on the total mass of system, mass ratio, and orbital period.

\begin{equation}\label{eqJ0}
J_0=\frac{q}{(1+q)^2} \sqrt[3] {\frac{G^2}{2\pi}M^5P}
\end{equation}

\begin{figure}
\centering
\includegraphics[scale=0.08]{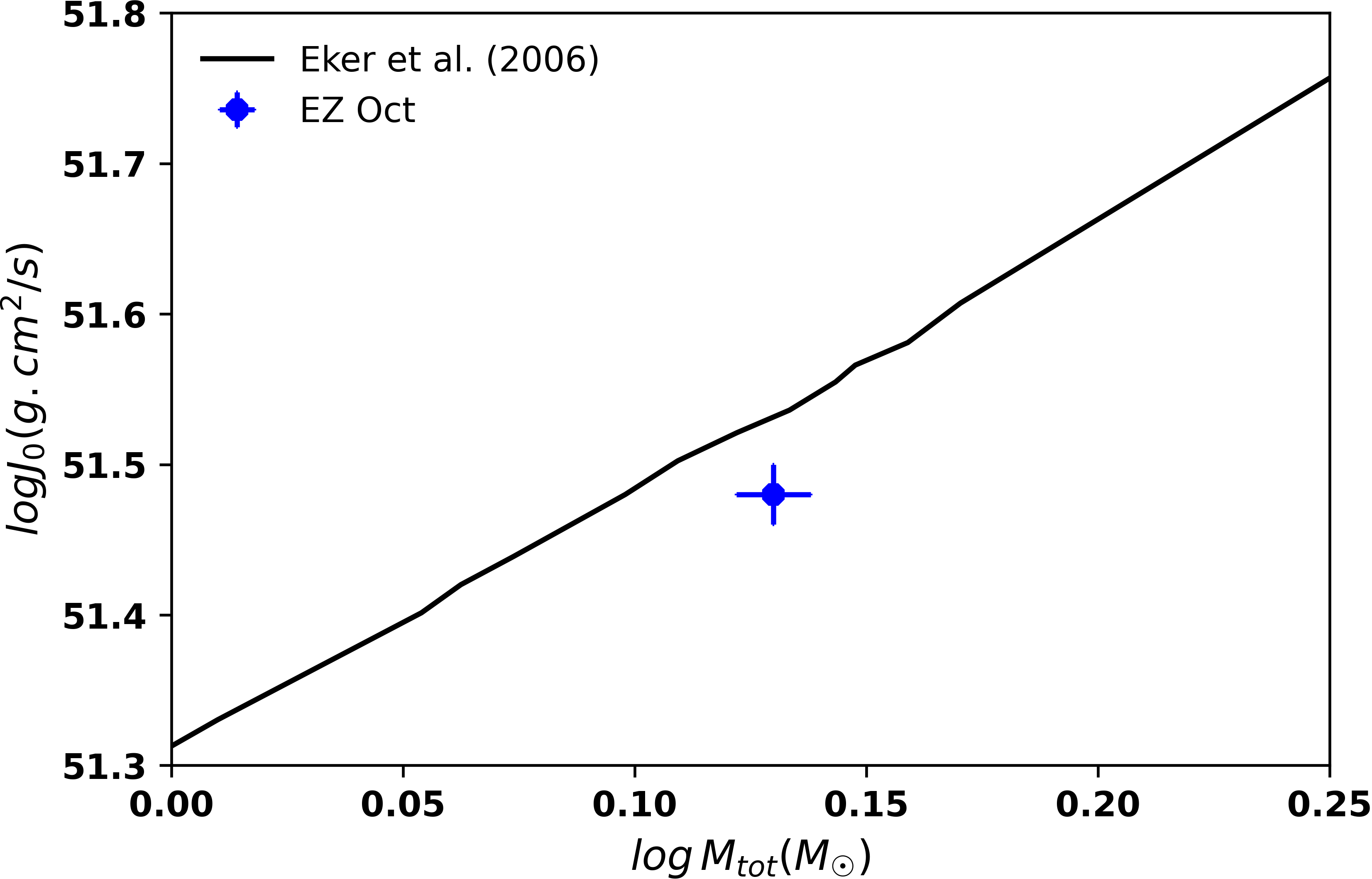}
\caption{Location of EZ Oct on the orbital angular momentum-total mass diagram.}
\label{Fig:J_0}
\end{figure}

The estimated orbital angular momentum, $\log J_0 \,(\mathrm{g\,cm^2\,s^{-1}})=51.48 \pm 0.01$, places EZ Oct securely within the region inhabited by contact binaries according to the criterion of \cite{2006MNRAS.373.1483E}, indicating that the system lies well beyond the detached/contact transition boundary (Figure \ref{Fig:J_0}). The photometric analysis yields a shallow degree of contact with a fill-out factor of $f = 0.106^{+0.008}_{-0.005}$, confirming an overcontact configuration rather than a marginal-contact state. Despite its shallow contact level, the system's geometric and dynamical properties consistently identify it as a stable member of the contact binary domain.

\begin{figure*}
    \centering
    \includegraphics[scale=0.085]{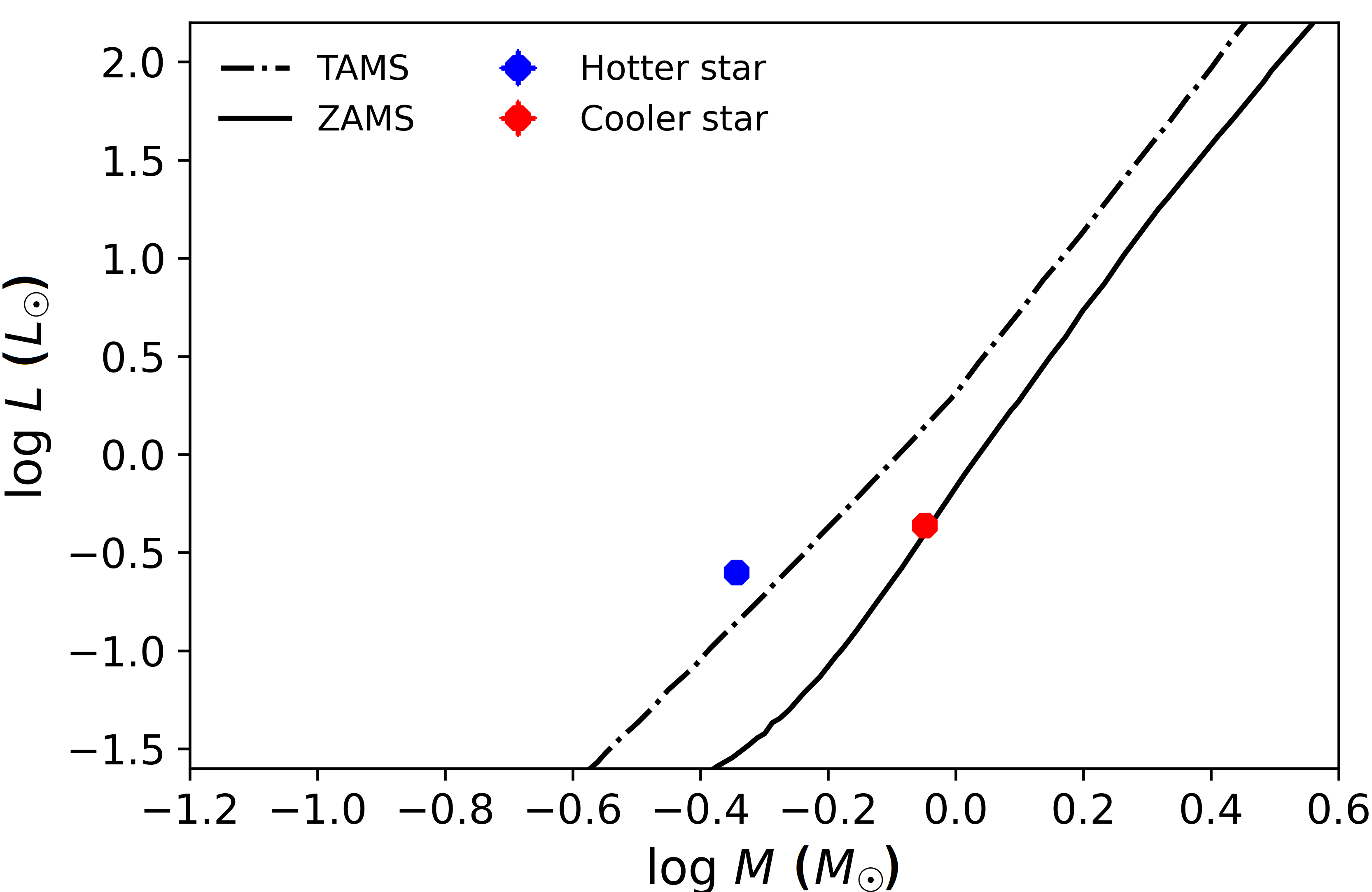}
    \includegraphics[scale=0.085]{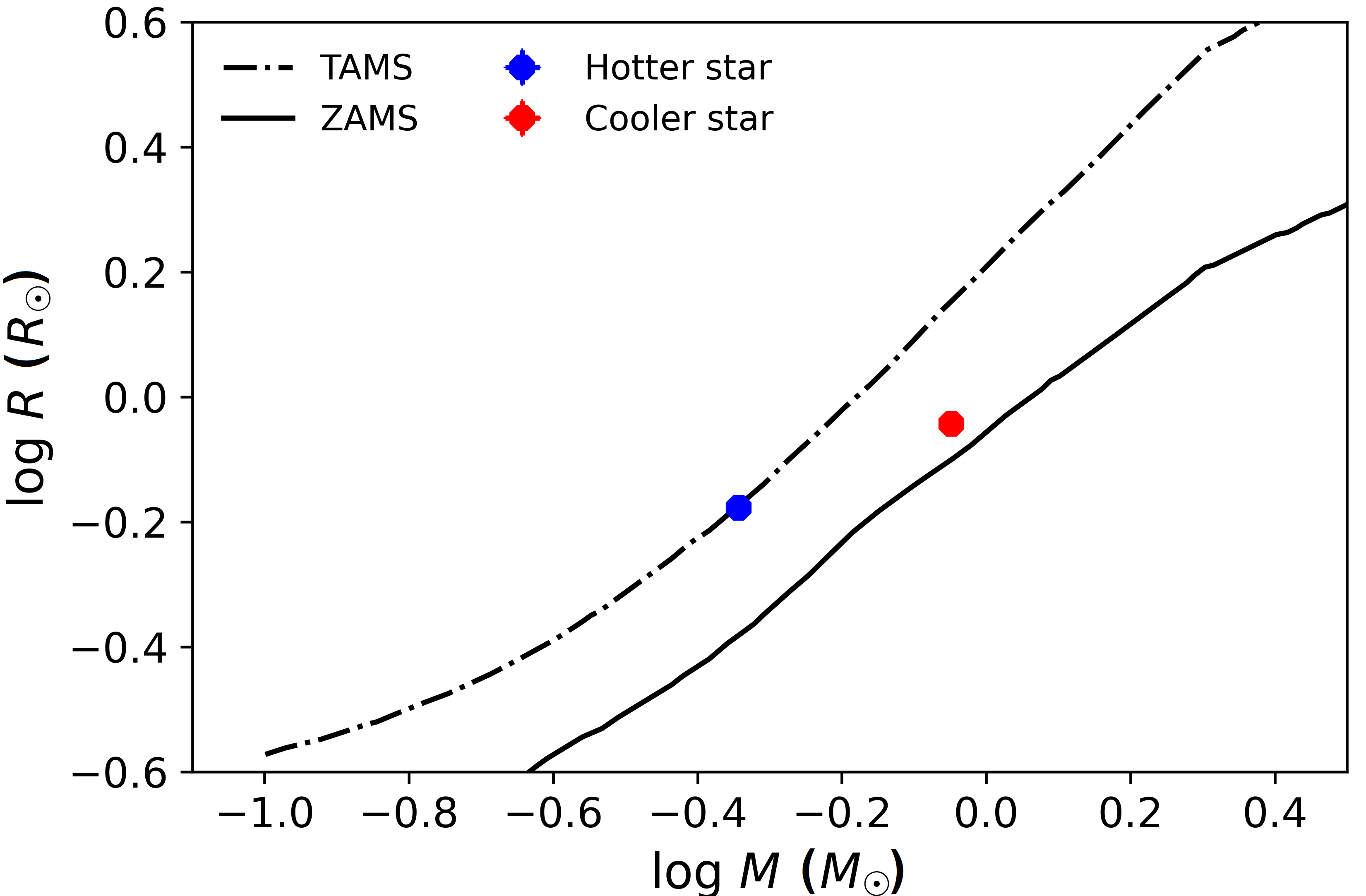}
    \caption{Logarithmic diagrams of the EZ Oct components showing mass versus radius and mass versus luminosity, relative to the ZAMS and TAMS lines.}
    \label{MLR}
\end{figure*}

F) The evolutionary status of EZ Oct is illustrated in the logarithmic Mass-Radius ($M$–$R$) and Mass-Luminosity ($M$–$L$) diagrams, plotted with the absolute parameters derived in this study (Table \ref{fundamental}, Figure \ref{MLR}). In these diagrams, the components of EZ Oct are shown with respect to the Zero-Age Main Sequence (ZAMS) and Terminal-Age Main Sequence (TAMS) tracks from \citet{2000AAS..141..371G}, indicating their evolutionary stage.

Light curve modeling combined with the derived absolute parameters indicates that the less massive component of EZ Oct is hotter than its companion. As illustrated in Figure \ref{MLR}, the lower-mass star is located near the TAMS, while the more massive component lies closer to the ZAMS, indicating that the two stars are at distinct evolutionary stages along the main sequence.

G) The light curve analysis, using parameters including mass ratio, fillout factor, and orbital inclination, confirms EZ Oct's contact binary nature. Using the classification scheme for A- and W-type contact binaries from \cite{1970VA.....12..217B}, and considering the derived component temperatures and masses, EZ Oct is identified as a W-subtype contact system. In such systems, the less massive star is hotter than its companion, which is consistent with the physical characteristics of EZ Oct.

H) The orbital Period–Luminosity ($P$–$L$) relationship in W UMa-type contact binaries has been extensively investigated following the study by \cite{1967MmRAS..70..111E}. \citet{1994PASP..106..462R} introduced a period–luminosity–color (PLC) relation based on 18 systems, which was subsequently refined by \cite{1997PASP..109.1340R} through the use of HIPPARCOS parallaxes and an expanded sample of 40 binaries. However, later studies (e.g., \citealt{2006MNRAS.368.1319R}; \citealt{2014MNRAS.443..432M}; \citealt{2016MNRAS.457.4323P}) yielded mixed results, often constrained by limited sample sizes or uncorrected period–color correlations. A more robust $P$–$L$ relation with approximately 10\% accuracy was presented by \cite{2016ApJ...832..138C}, using a sample of 66 contact binaries. Building on this, \cite{2018ApJ...859..140C} employed 183 close systems with precise parallaxes from the Tycho–Gaia Astrometric Solution (TGAS) and included both optical and mid-infrared photometry. Then, \cite{2021ApJS..254...10L} analyzed 210 W UMa-type systems with orbital periods below 0.5 days, deriving distinct $P$–$L$ relations for the primary and secondary components (Equation \ref{eq:L2021}). More recently, \cite{2024RAA....24a5002P} investigated the $P$–$L$ relationship using a sample of 118 contact systems, with luminosities estimated via a method based on Gaia DR3 parallaxes. They provided revised $P$–$L$ relations separately for the primary and secondary components (Equation \ref{eq:Poro2024}).

\begin{equation}
\left.
\begin{aligned}
\log L_1=(13.98\pm0.75)P-(3.04\pm0.27) \\
\log L_2=(3.66\pm0.26)P-(0.69\pm0.09)
\end{aligned}
\right\} 
\quad \text{\citet{2021ApJS..254...10L}}
\label{eq:L2021}
\end{equation}

\begin{equation}
\left.
\begin{aligned}
\log L_1 &= (4.15 \pm 0.11)\, \log P + (1.90 \pm 0.13) \\
\log L_2 &= (2.82 \pm 0.12)\, \log P + (1.08 \pm 0.15)
\end{aligned}
\right\} 
\quad \text{\citet{2024RAA....24a5002P}}
\label{eq:Poro2024}
\end{equation}

\begin{figure*}
    \centering
    \includegraphics[scale=0.2]{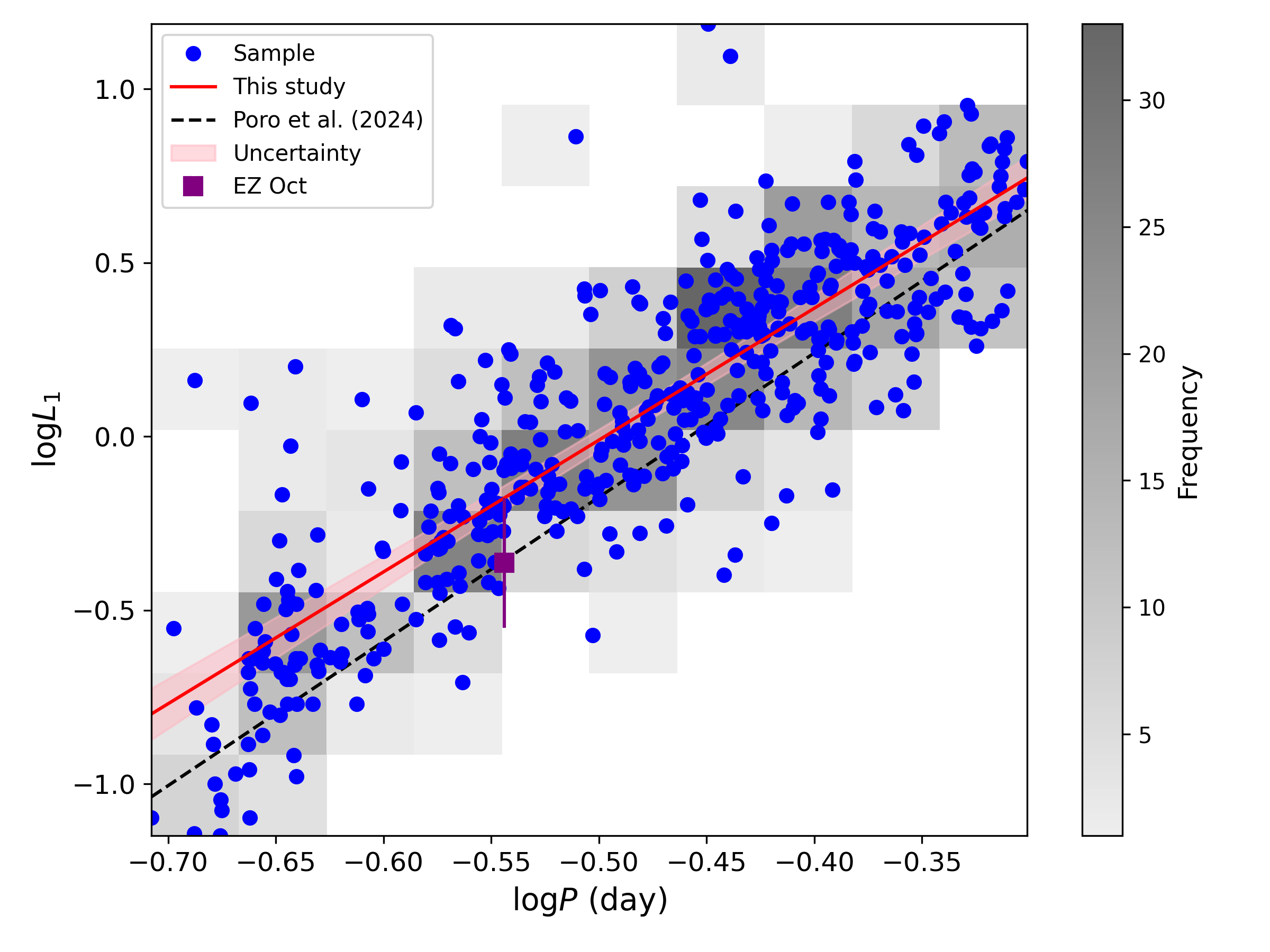}
    \includegraphics[scale=0.2]{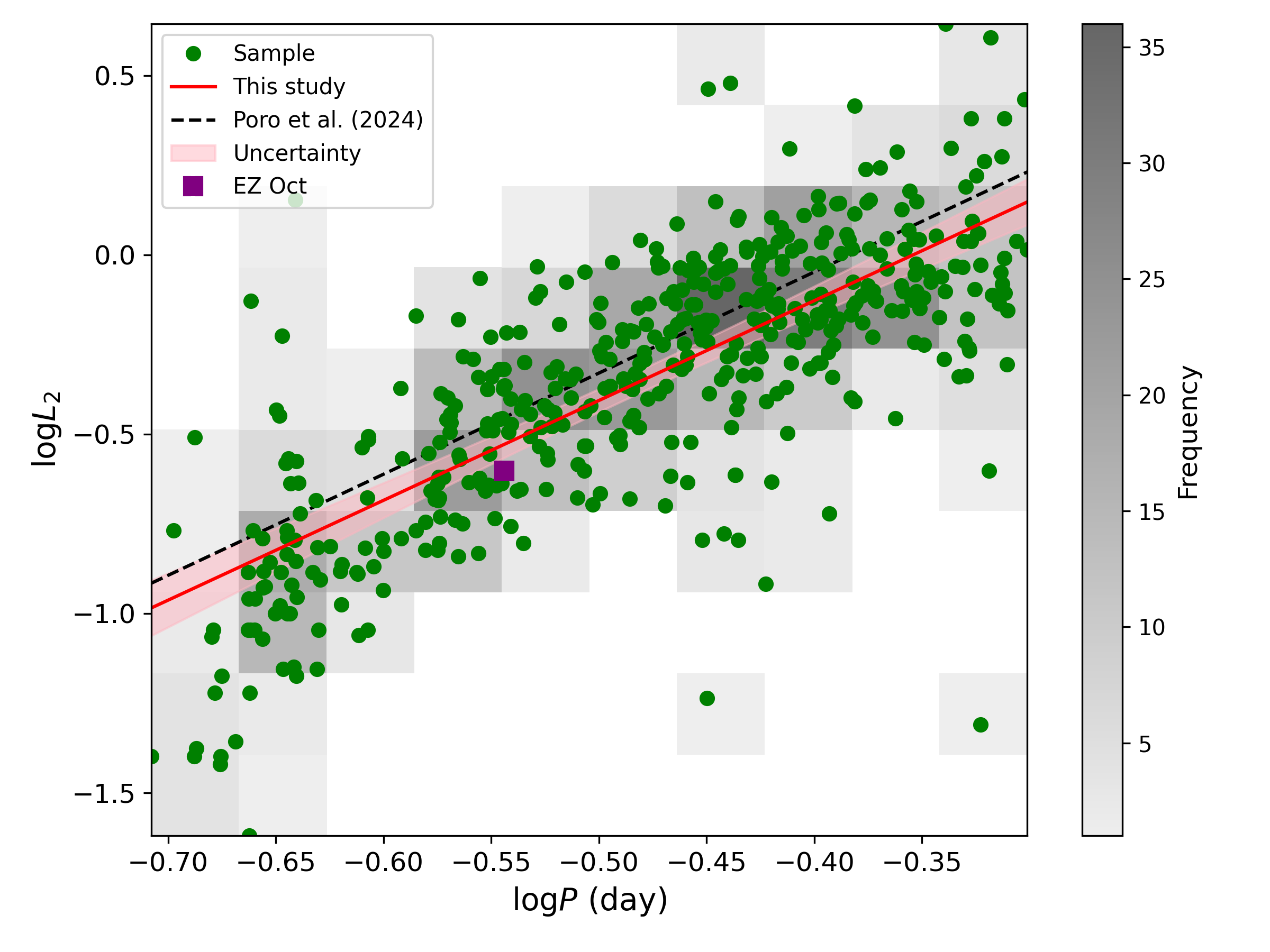}
    \caption{2D plots illustrating how orbital period correlates with luminosity for stars in contact binaries.}
    \label{PL}
\end{figure*}

This investigation revisits the empirical $P$–$L$ parameter relationship using an expanded dataset from the \cite{2025MNRAS.538.1427P} study, exceeding the sample sizes of previous studies. Our sample includes 461 contact systems with orbital periods shorter than 0.5 days. A linear fit was applied to the data, as shown in Figure \ref{PL}, to examine the overall trend of the $P$–$L$ relationship. A heatmap was also overlaid on the diagrams to illustrate the density distribution of the systems across the parameter space. The revised $P$–$L_1$ and $P$–$L_2$ relations are presented as follows (Equations \ref{P-L1}, \ref{P-L2}):

\begin{equation}\label{P-L1}
log L_1 = (3.79 \pm 0.10) \times log P + (1.89 \pm 0.05),
\end{equation}

\begin{equation}\label{P-L2}
log L_2 = (2.78 \pm 0.11) \times log P + (0.99 \pm 0.05).
\end{equation}

The position of the EZ Oct system is marked on the diagrams in Figure \ref{PL}. In the upper panel of Figure \ref{PL} ($P$–$L_1$), the position of EZ Oct exhibits a slight deviation from the newly derived linear fit, whereas in the lower panel ($P$–$L_2$), the target aligns well with the linear fit. The position of the EZ Oct system and the sample systems is not necessarily expected to coincide with the newly derived fit and its uncertainties. As shown in Figure \ref{PL}, other systems from the literature are located near EZ Oct, while some systems lie outside the uncertainty range of the linear fit. Compared to the study by \cite{2024RAA....24a5002P}, which examined these relations using 118 systems, the current study employs approximately fourfold more sample systems. Naturally, the number of systems used in each empirical relation can significantly affect the results. The empirical relations in binary systems are employed solely to examine trends and to study the mutual influence among parameters.

I) We investigated the relationship between the mass ratio and the orbital inclination parameters for our target system EZ Oct together with a comparison dataset of 809 contact binaries compiled from \cite{2025MNRAS.538.1427P}. The binaries in this sample span orbital inclinations between $40^\circ$ and $90^\circ$, while the reciprocal mass ratios are confined to the range $0<q \leq 1$.

\begin{figure*}
\centering
\includegraphics[width=0.99\textwidth]{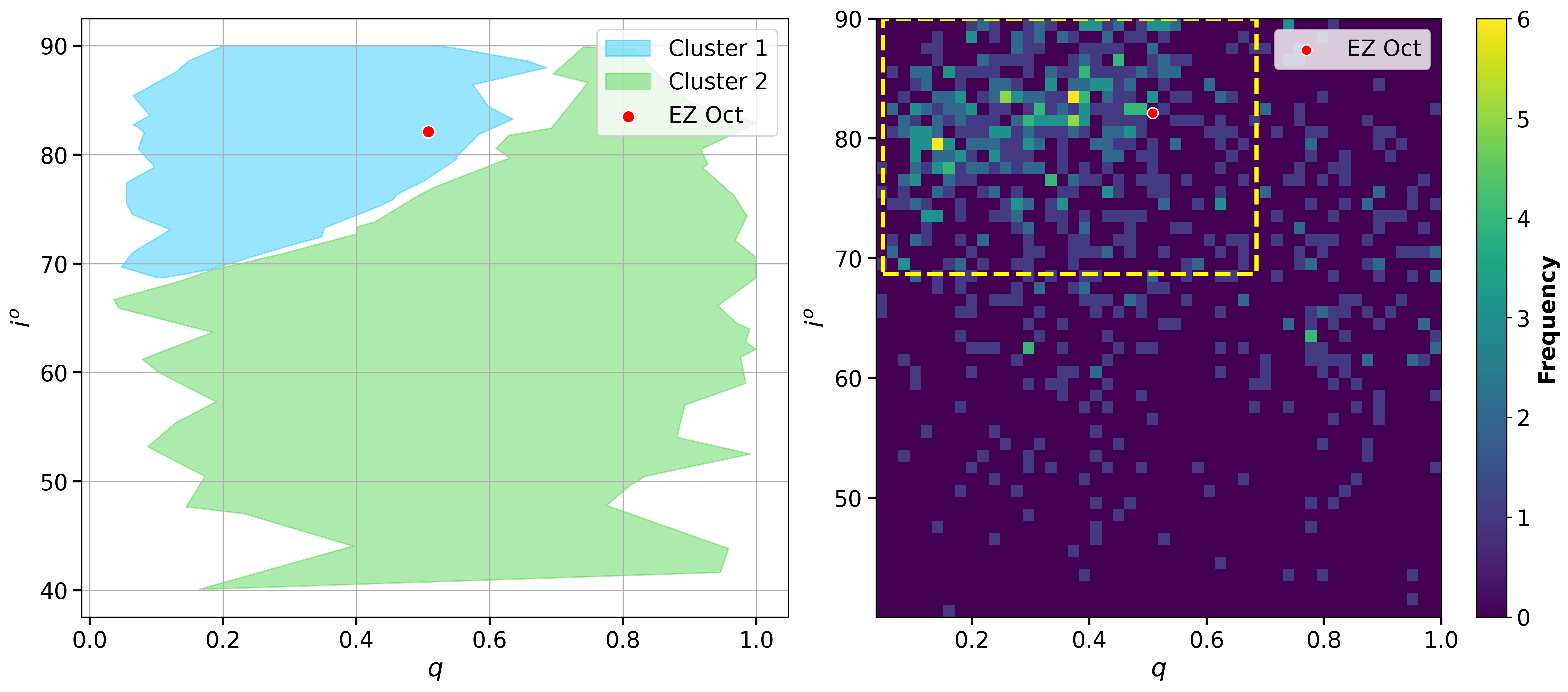}
\caption{hl{Density distribution of the contact binary sample in the $q$-$i$ space. Left panel: GMM-based clustering with alpha-shape boundaries. Right panel: 2D histogram heatmap of the same data. The red marker in both panels indicates EZ Oct, and the dashed yellow rectangle in the right panel corresponds to the region occupied by Cluster 1.}}
\label{q-i}
\end{figure*}

Gaussian Mixture Model (GMM; \citealt{2011JMLR...12.2825P}) clustering was applied to the data to identify distinct groups based on density structures in the $q$-$i$ space (left panel of Figure \ref{q-i}). Cluster 1 corresponds to the densest region of the distribution, and the target system EZ Oct (marked in red) is located within this cluster. A two-dimensional density heatmap was then constructed (right panel of Figure \ref{q-i}) to examine the frequency distribution of systems across the $q$-$i$ parameter space. Brighter regions correspond to higher concentrations of data points and darker areas to lower densities, with a binning resolution of $50 \times 50$ adopted to balance global structures and local variations. The range of Cluster 1 is highlighted by a dashed yellow rectangle, enabling a focused analysis of the densest subset of systems.

The Pearson correlation coefficient (PCC) between $q$ and $i$ was calculated over the entire dataset of 809 binaries, resulting in a moderate value of -0.254. Within the rectangle corresponding to Cluster 1 (right panel of Figure \ref{q-i}), the PCC is 0.123, and approximately 70\% of the dataset falls inside this region. Although the global PCC indicates a weak inverse correlation, the distribution in Figure \ref{q-i} shows that binaries within Cluster 1 exhibit a slight positive correlation, with orbital inclination tending to increase with mass ratio. This pattern reflects that the overall dataset appears negatively correlated mainly due to the distribution of binaries outside the densest cluster, whereas the majority of systems within Cluster 1 follow a locally positive trend. This combined analysis of the heatmap, clustering, and one-dimensional distributions illustrates that the overall dataset shows a weak inverse correlation, whereas the behavior within the densest cluster clearly exhibits a positive trend with increasing mass ratio.

\vspace{0.6cm}
\section*{Data Availability}
The observational data and extracted minima are accessible in the online supplementary files.

\vspace{0.6cm}
\section*{Acknowledgments}
We sincerely thank the referee for their constructive comments, which significantly improved the quality of this manuscript. Observations, analyses, and writing for this study were carried out within the BSN project framework. Data from the ESA Gaia mission\footnote{\url{http://www.cosmos.esa.int/gaia}} were used, and IRAF software, maintained by the National Optical Observatories and managed by the Association of Universities for Research in Astronomy under a cooperative agreement with the NSF, was employed for data reduction. We also acknowledge the use of TESS data from the NASA mission, obtained through MAST for this work. We appreciate Ebubekir Atsız's assistance in preparing the O–C diagram.

\vspace{0.6cm}
\bibliography{References}{}
\bibliographystyle{aasjournal}

\end{document}